# DIAGNOSTICS FOR DESIREE

S. Das and A. Källberg[#], Manne Siegbahn Laboratory, Stockholm University, Frescativägen 26, S - 11418 Stockholm, Sweden

*Abstract*

The beam diagnostics system for the Double ElectroStatic Ion Ring ExpEriment (DESIREE) project under construction is briefly described. A system which can operate over a wide range of beam intensities and energies for the DESIREE has been built and tested using the CRYRING facility at MSL. Spatial resolution up to 2 mm is achieved.

## INTRODUCTION

Single electrostatic ion storage rings have attracted considerable attention in recent years in atomic and molecular physics because of their several advantages over magnetic storage rings [1-6]. Motivated by the success of electrostatic ion storage rings and the possibility of performing merged-beams experiments with positive and negative ions, a Double ElectroStatic Ion Ring ExpEriment (DESIREE) has been under construction at Stockholm University and will soon be ready for experiments [7-9].

In this report we briefly describe the different elements to be used in the beam diagnostics system for DESIREE. A diagnostic system for radioactive beam experiment (REX) using the beam profile monitoring system (BPMS) [10,11] as part of DESIREE beam line has been built and tested for spatial resolution using the existing CRYRING facility at MSL [9]. A resolution of 2 mm has been achieved.

## DESIREE

A schematic layout of the DESIREE is shown in Fig.1. It consists of two rings of same circumference of 9.2 m and a common straight section of length 1 m for the merged-beams experiments with ions of opposite charges. DESIREE will be operated at both room and cryogenic temperatures (~10-20 K) under ultra high vacuum (~$10^{-11}$ mbar) environment. In addition to the merged-beam experiments, DESIREE will be also used for single ring experiments.

The basic diagnostics components that will be used are Faraday cups (FC), electrostatic pickups (PU), Microchannel plates (MCP), and scrapers. FC will be used to measure the beam current before and after the injection of the beam in the storage rings, PU to measure the transverse beam position in the rings. Position sensitive MCP detectors will be used to detect the charged reaction products and the neutral particles that are occurred from the merged-beam experiments. Beam scrapers will be used in the straight merger section to attain the maximum beam size both horizontally and vertically. It should be noted, however, that the properties of the detectors such as MCP change very rapidly with temperature and at cryogenic temperature very little data exist until now. Therefore, the development of the beam diagnostic system for the DESIREE project is a challenging one.

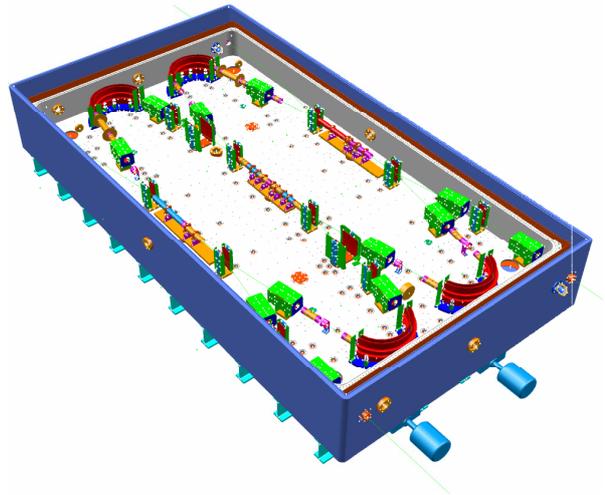

Figure 1: Schematic layout of DESIREE.

## BEAM DIAGNOSTIC SYSTEM FOR REX

A diagnostic system based on the observation of low energy (~10 eV) secondary electrons (SE) produced by a beam, striking a metallic foil has been built to monitor and to cover the wide range of beam intensities and energies [10,11].

The system consists of (i) a Faraday cup (FC) to measure the beam current, (ii) a collimator with circular apertures of different diameters to measure the spatial resolution of the system (Fig. 2), (iii) a beam profile monitoring system (BPMS), and (iv) a control unit (PC). The BPMS, in turn, consists of (a) an aluminim (Al) plate or foil, (b) a grid placed in front of the Al foil to accelerate the SE, (c) position sensitive MCP, (d) fluorescent screen (F.S), and (e) a CCD camera to capture the images (Fig. 3). The spatial resolution of the system depends on the voltage applied on the Al plate. The amplification of the MCP can be controlled by the applied voltage.

The collimator contains a set of circular holes of different diameters ($\Phi$) and separations ($d$) between them as shown in Fig. 2. The collimator cuts out from the beam areas equal to the holes with separation $d$ mm between the beams centers and creates well separated (distinguishable) narrow beams of approximately same intensity close to each other.

[#] kallberg@msl.se

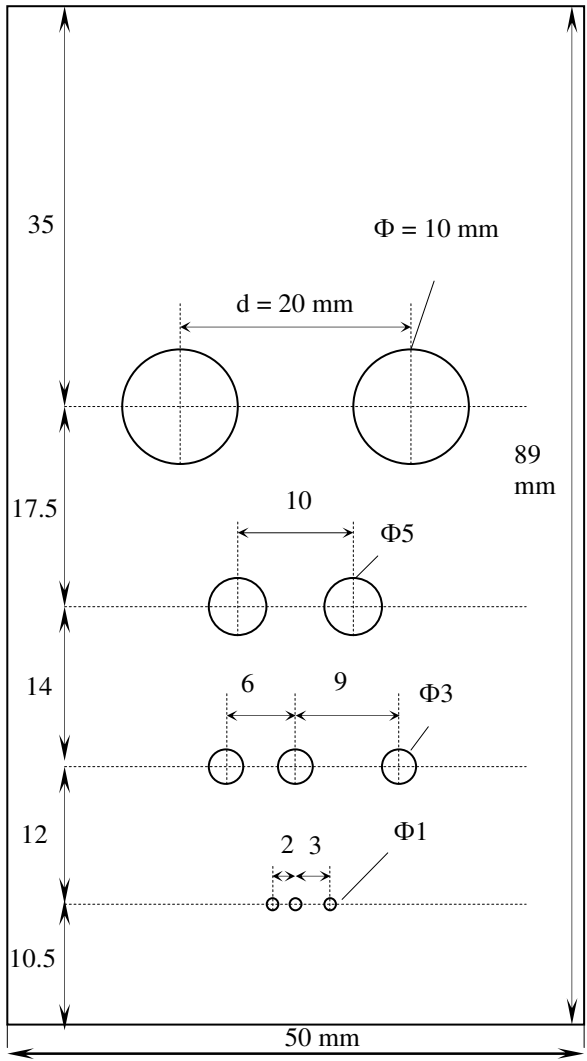

Figure 2: Collimator with circular holes of different diameters. Φ is the diameter of hole. d is the separation between the adjacent holes.

The three-dimensional (3-D) view of the BPMS is shown in Fig. 3. The schematic diagram and experimental setup of the beam diagnostic system are shown in Figs. 4 and 5, respectively. The collimator and the Al foil attached with the grid can be inserted and taken out of the beam path by a pneumatic feed-through. The 10 keV proton beam from the CRYRING source passes through the collimator, strikes the Al plate, and knocks off low energy (~ 10 eV) secondary electrons (SE). These SE are then accelerated by a homogeneous electric field applied between the Al plate and the grid that are kept at negative and ground potentials, respectively. The accelerated SE from the grid then travel the field free region to the position sensitive electron multiplier, micro channel plate (MCP). The amplified electrons from the MCP then hit the fluorescent screen (F.S) and produce flashes of light. The F.S is kept at ground potential. The light is captured by the computer controlled CCD camera [12].

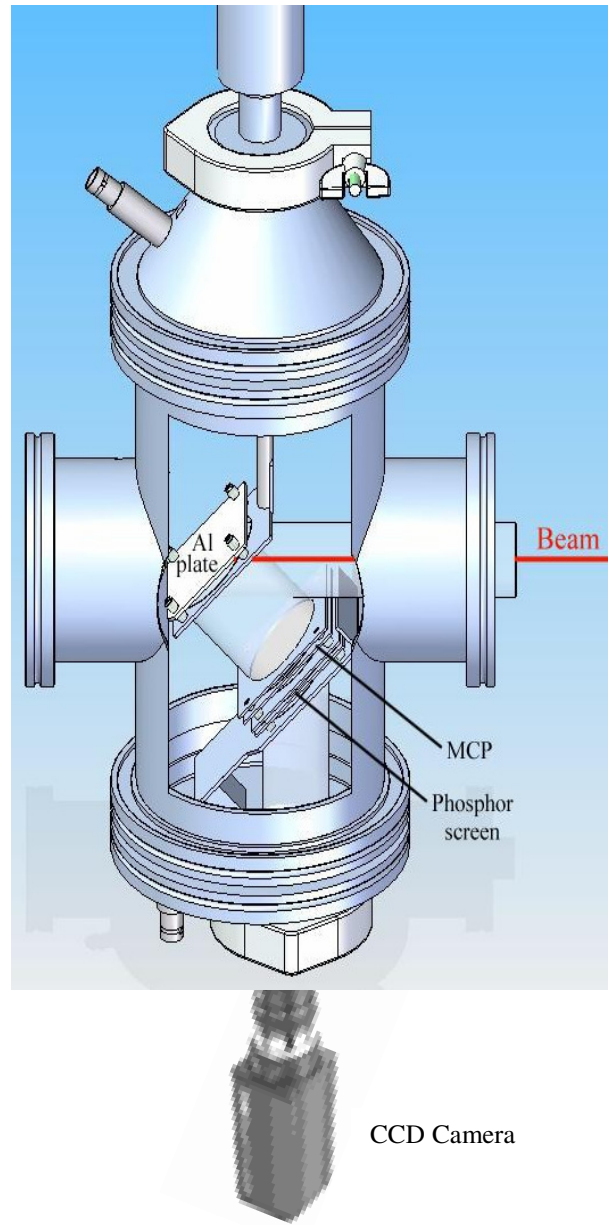

Figure 3: Three dimensional (3-D) view of the BPMS.

The CCD images of the collimator holes for fixed Al plate and MCP voltage recorded on the PC are shown in Fig. 6. It can be easily form the Fig. 6 that the collimator has created well separated beams. The figure shows that a spatial resolution of 2 mm can be achieved. The rest of the article is focused on the two holes of diameter 1 mm which are separated by 2 mm (see Fig. 2). The resolution of the system was tested for different Al plate and MCP voltages and the results are shown in Fig. 7.

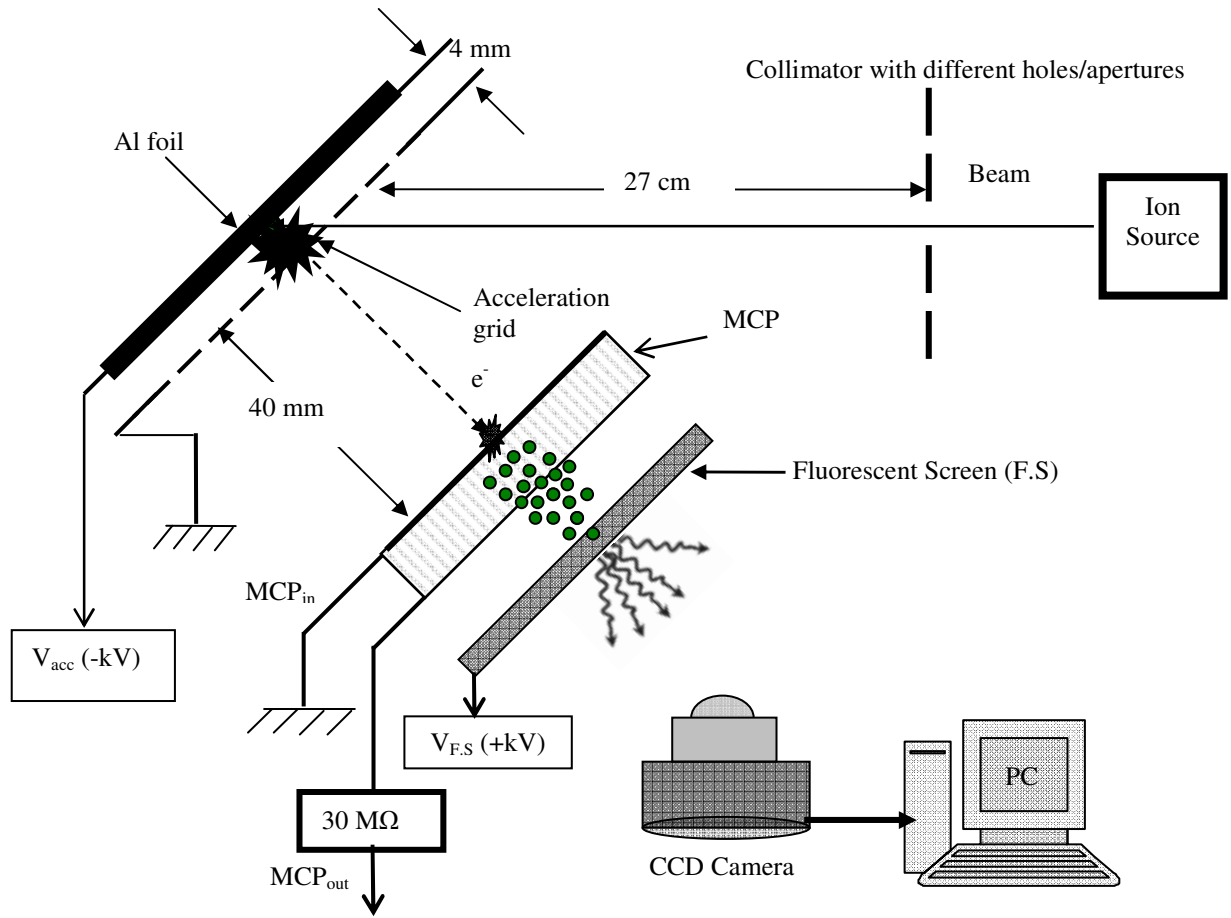

Figure 4: Schematic diagram of the beam diagnostic setup.

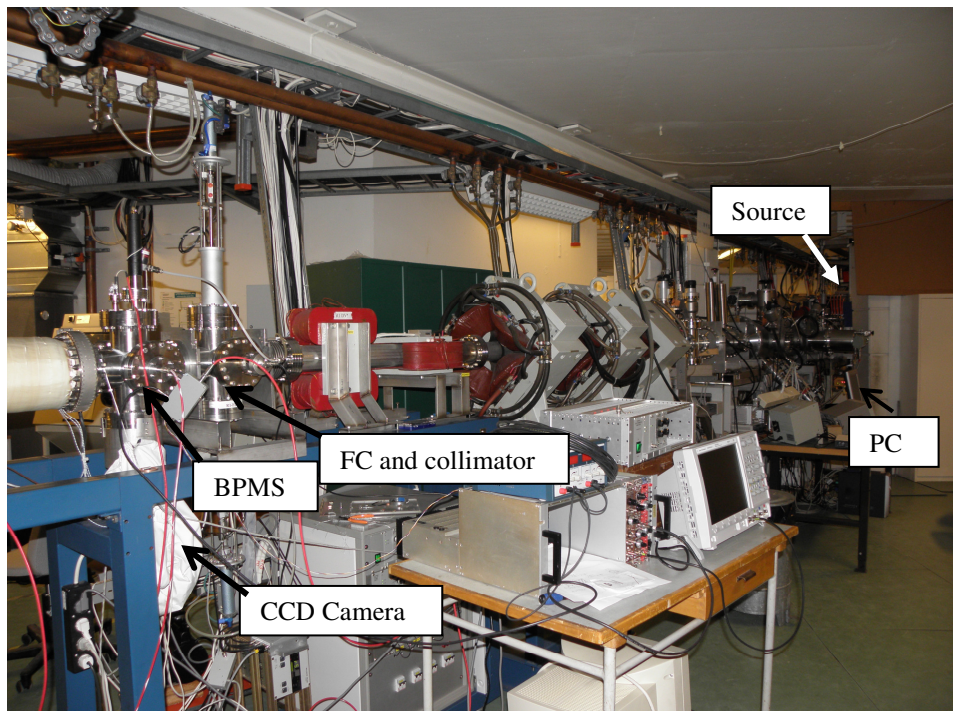

Figure 5: Experimental setup at the CRYRING facility at MSL, SU.

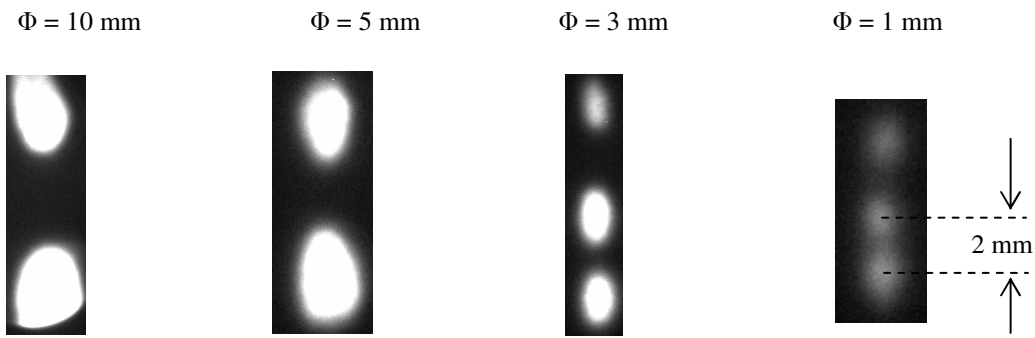

Figure 6: CCD images of the light produced in the fluorescent screen (F.S) for fixed Al plate ($V_{acc}$ = - 5 kV) and MCP ($V_{MCP}$ = 1300 V) voltage. The beam passed through the collimator and generated SE from the Al plate. Well separated beam spots are clearly seen on the screen showing a spatial resolution of better than 2 mm of the diagnostic system.

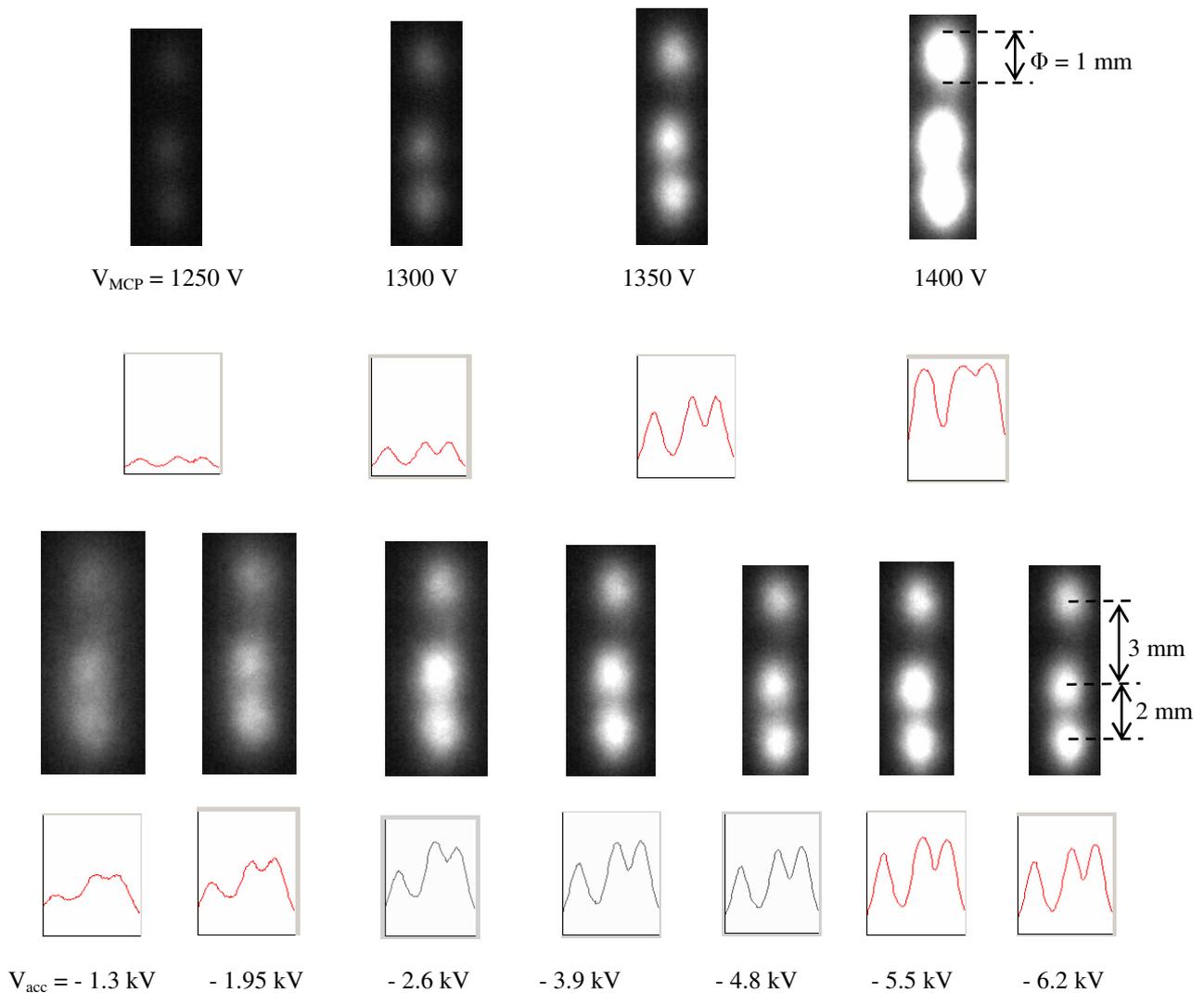

Figure 7: CCD images of the light produced in the F.S. The intensity distribution is shown as a function of $V_{MCP}$ ($V_{acc}$ = - 5 keV) (upper panel) and $V_{acc}$ ($V_{MCP}$ = 1370 V) (lower panel). The vertical beam profiles captured by the CCD camera are also shown.

Moreover, the change of resolution of the system as a function of Al plate voltage was also measured from the beam profiles recorded on the CCD camera (see lower panel of Fig. 7). For this, the length of the dip ($\Delta$) between the two adjacent peaks was calculated as $\Delta = \left(\frac{a+b}{2} - c\right)$. Here, *a*, *b*, and *c* are the heights of two adjacent peaks, and the dip between them, respectively, as shown in Fig. 8. The ratio $\delta = \frac{\Delta}{\left[\frac{(a+b)}{2}\right]}$ was then plotted as a function of Al plate voltage (Fig. 9). It can be seen form the Fig. 9 that the resolution of the system increases with increasing plate voltage.

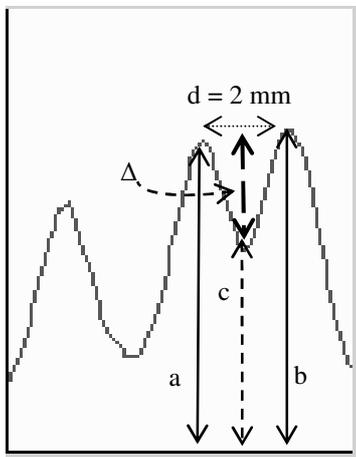

Figure 8: Vertical beam profile captured on the camera. Two apertures of diameter 1 mm, separated by 2 mm are considered. a, b, and c are the heights of the two adjacent peaks, and the dip between them, respectively. (a+b)/2 is average of the heights of two adjacent peaks.

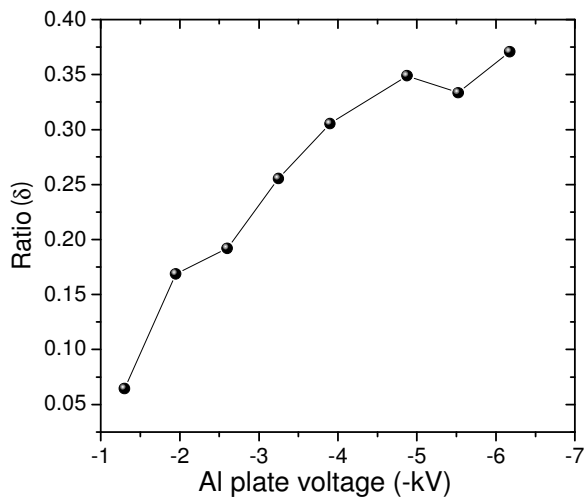

Figure 9: Plot of ratio as a function of Al plate voltage.

## CONCLUSIONS

In summary, we have described briefly the diagnostics system for the DESIREE project. A system has been built using the CRYRING facility at the Manne Siegbahn Laboratory, Stockholm University as part of the DESIREE beam line diagnostics. A spatial resolution of better than 2 mm has been achieved. However, a more detailed and systematic investigation to improve the resolution of the system further is required.

## ACKNOWLEDGEMENT

One of the authors (S. Das) acknowledges the financial support received from the Marie Curie Fellowship.

## REFERENCES

[1] S.P. Møller, Nucl. Inst. Methods. Phys. Res. Sec. A 394 (1997) 281.
[2] T. Tanabe *et al.*, Nucl. Inst. Methods. Phys. Res. Sec. A 482 (2002) 595.
[3] T. Tanabe et al., Nucl. Inst. Methods. Phys. Res. Sec. A 496 (2003) 233.
[4] T. Tanabe *et al.*, Nucl. Inst. Methods. Phys. Res. Sec. A 532 (2004) 105.
[5] S.P. Møller *et al.*, Proc. EPAC 2000, Austria, p. 788.
[6] C.P. Welsch *et al.*, Hyperfine Interactions 146/147 (2003) 253.
[7] K. –G. Rensfelt *et al.*, Proc. EPAC 2004, Switzerland, p. 1425.
[8] P. Löfgren et al., Proc. EPAC 2006, Scotland, p. 252.
[9] Manne Siegbahn Laboratory (www.msl.se), and Atomic Physics, Stockholm University http://www.atom.physto.se/Cederquist/desiree_web_hc.html.
[10] K. Kruglov *et al.*, Nucl. Inst. Methods. Phys. Res. Sec. A 441 (2000) 595.
[11] K. Kruglov *et al.*, Nucl. Inst. Methods. Phys. Res. Sec. A 701 (2002) 193c.
[12] Matrix Vision (http://www.matrix-vision.com).